\documentclass[sigconf]{acmart}

\usepackage{booktabs} 
\usepackage{wrapfig}
\usepackage{dblfloatfix}
\usepackage{hyperref}
\usepackage{balance}
\usepackage{colortbl}
\usepackage{mathtools}
\usepackage{cleveref}
\usepackage[shortlabels]{enumitem}
\usepackage[most]{tcolorbox}

\usepackage{picture}

\usepackage{pifont}

\newcommand{\bi}{\begin{itemize}[leftmargin=0.4cm]}
	\newcommand{\ei}{\end{itemize}}
\newcommand{\be}{\begin{enumerate}}
	\newcommand{\ee}{\end{enumerate}}

\makeatletter
\let\th@plain\relax
\makeatother

\definecolor{lightgray}{gray}{0.8}
\definecolor{darkgray}{gray}{0.6}
\definecolor{Gray}{gray}{0.85}
\definecolor{Blue}{RGB}{0,29,193}
\usepackage{tikz}
\usepackage{framed}
\usepackage[framed]{ntheorem}
\usetikzlibrary{shadows}

\theoremclass{Lesson}
\theoremstyle{break}

\tikzstyle{thmbox} = [rectangle, rounded corners, draw=black,
fill=Gray!40]

\newshadedtheorem{lesson}{Result}

\usepackage{listings}
\definecolor{MyDarkBlue}{rgb}{0,0.08,0.45} 
\setcopyright{rightsretained}

\acmDOI{10.475/123_4}

\acmISBN{123-4567-24-567/08/06}

\acmConference[MSR'18]{Mining Software Repositories}{May 2018}{Gothenburg, Sweden} 
\acmYear{2018}
\copyrightyear{2018}

\acmArticle{4}
\acmPrice{15.00}

\begin{document}
\title{ 500+ Times Faster Than Deep Learning} 
\subtitle{(A Case Study Exploring Faster   Methods for Text Mining StackOverflow)}

\author{Suvodeep Majumder, Nikhila Balaji, Katie Brey, Wei Fu, Tim Menzies}  
\affiliation{%
  \institution{Computer Science, NC State, USA}
}
\email{smajumd3,nbalaji,kebrey, wfu@ncsu.edu; tim@menzies.us}

\renewcommand{\shortauthors}{S. Majumder et al.}


\begin{abstract}
 Deep learning methods are 
useful for  high-dimensional data and are becoming widely used in many areas of software engineering. 
Deep learners utilizes extensive computational power and can take a  long time to train-- making it difficult to widely validate and repeat and improve their results.
Further, they are not the best solution in all domains. For example, recent results
show that for finding related   Stack Overflow posts, a tuned SVM performs similarly to a deep learner, but is significantly faster to train. 

This paper extends that recent result by   clustering the dataset, then    tuning very learners within each cluster. This approach is over  500 times faster than  deep learning (and over 900 times faster if we use all the cores on a standard laptop computer). Significantly, this faster approach generates classifiers   nearly as good (within 2\% F1 Score) as the much slower deep learning method.
Hence we recommend this faster methods since it is much easier to reproduce and utilizes far fewer CPU resources. 

More generally, we recommend that before researchers release research results, that they compare their supposedly sophisticated methods against simpler alternatives
(e.g applying simpler learners to build local models).
\end{abstract}

%
%



\maketitle

 \section{Introduction}
    \label{sect:intro}
   Recently, deep learning methods like   convolutional neural networks (CNN) have
   become a   popular choice for text mining in SE. Such   deep learning works well with high dimensional data~\cite{mou2016convolutional} but are very expensive in terms of time and required CPU.

  \begin{table}[!b]
        \centering
        \small
        \begin{tabular}{c|c|p{4.5cm}}
            \rowcolor{gray} \textcolor{white}{\textbf{Type}} & \textcolor{white}{\textbf{Abbreviation}} & \textcolor{white}{\textbf{Learner Description}}\\
            \hline
            G & SVM & Support Vector Machine \\ 
            \rowcolor{lightgray} G & KNN & K-nearest Neighbors \\ 
            G & DE\_KNN & K-nearest Neighbors tuned using differential evolution (DE)\\ 
            \rowcolor{lightgray} G & DE\_SVM & Support Vector Machines Tuned using differential evolution \\
            G & CNN & Convolution Neural Network \\
            \rowcolor{lightgray}L & KMeans\_KNN & Cluster + multiple KNNs \\ 
            L & KMeans\_DE\_KNN & Cluster + multiple KNNs tuned via DE \\ 
            \rowcolor{lightgray}L & KMenas\_SVM & Cluster + multiple SVMs \\ 
            L & KMeans\_DE\_SVM & Cluster multiple SVMs tuned via DE \\ 
        \end{tabular}\vspace{3mm}
        \caption{Combinations of learning algorithms.
        L= ``local models''  and G= ``global models'' which denotes learning from clusters or all
        of the data (respectively).}
        \label{tab:learners}
    \end{table}

   In response to the computational problems of
   deep learners, researchers have tried alternate methods.
   For example,
   Fu et al. ~\cite{fu2017easy} recently revisited
   a study by Xu et al. ~\cite{xu2016predicting}   on finding the semantic relatedness of Stack Overflow posts.
   In that study, a single Stack Overflow question along with its complete answer was called a ``knowledge unit'' (KU). If any two KUs are semantically related, they are considered as {\em linkable} knowledge. Otherwise, they are considered {\em isolated}.   When Xu et al. applied CNN ~\cite{hubel1959receptive} (a specific kind of deep learner), that took 14 hours to train. Fu et al.
   showed
   that this computational cost was avoidable
   by replacing a more complex learner (CNN)
   with a simpler technique augmented by some
   hyperparameter optimization.
   Specifically, Fu et al. shows that   support vector machine (SVM) tuned via differential evolution (DE) could  perform as well as CNN, while
   training   84 times  faster.  
    
   This paper further extends the Fu et al. Using very simple widely used data mining method (KMeans), we can train even faster that Fu et al. and 
   500 times faster than  deep learning (and over 900 times faster if we use all the cores on a standard laptop computer).
   The core to our approach is  (1)~{\em building multiple local models} then (2)~{\em tuning per local model}.
This paper evaluates this divide and conquer approach by:
    \begin{enumerate}
        \item  Exploring the Xu et al. task using SVM and K-nearest-neighbor (KNN) classifiers;
        \item Repeating step 1 using  hyperparameter tuning--
        specifically,  differential Evolution (DE)--
        to select control parameters for those learners;
        \item Repeats steps 1 and 2 using 
        {\em local modeling};
        i.e. clustering the data then apply tuning and learning to each cluster; 
      
        \item Evaluating these local models in terms of both their training time and performance
    \end{enumerate}
 Table~\ref{tab:learners} lists all the learners explored here.
 
  \newpage\noindent
  This approach lets us ask and answer the following questions:
    \begin{itemize}
        \item 
            \textbf{RQ1}:  Can we reproduce Fu et al.'s results for tuning SVM with differential evolution (DE)?
            \begin{lesson}
                Our DE with SVM perform no worse than Fu et al.
            \end{lesson}
        \item 
            \textbf{RQ2}:    How do the local models compare with global models in both tuned and untuned versions in terms model training time?
            \begin{lesson}
                Local models perform comparably to their global model counterparts, but are 570 times faster in model training time.
            \end{lesson}
            (To be precise,  that   570 figure comes from running on a single core. If we distribute the execution cross the eight cores of a standard laptop computer, our training times
            become 965 times faster.)
        \item 
            \textbf{RQ3}:   How does the performance of local models compare with global models and state-of-the-art deep learner when used with SVM and KNN?
            \begin{lesson}  
                Local models performance very nearly as well (within  2\% F1 Score) as their global counterpart and the state-of-the-art deep Learner.
            \end{lesson}
    \end{itemize}
    Based on these experiments and discoveries, our contribution and outcome from the paper are:
    \begin{itemize}
     \item  A dramatically faster solution to the Stack Overflow text mining task first
     presented by Xu et al. This  new method
     runs three orders of magnitude faster than prior work.
        \item Support for ``not everything needs deep learning'';  i.e. sometimes, applying
        deep learning to a problem may not be the
        best approach.
        \item Support for  a simplicity-first approach; i.e.
         simple method like KMeans\_DE\_SVM   can performs as good some of the state of the art models but with a (much) faster training time.
        \item Support for local modeling. Such local
        models   can significantly reduce training time by  clustering data then restricting learning to  on each cluster.  
        \item  A reproduction package - which can be used to reproduce, improve or refute our results\footnote{URL blinded for review.}.

    \end{itemize}
    The rest of the paper is organized into the following sections 
    Section~\ref{sect: Background and Motivation} provides background information  that directly relates to our research questions, in addition to laying out the motivation behind our work. In 
    Section~\ref{sect: Experimental Design} a detailed description of our experimental setup and data, along with our performance criteria for evaluation is presented. It is followed by 
    Section~\ref{sect:Result} the results of the experiments and answers to our research questions are detailed.
    Section~\ref{sect:THREATS TO VALIDITY} discusses threats to validity. Finally 
    Section~\ref{sect:CONCLUSION} concludes the paper with implications and scope for future work.
\section{Background  }
\label{sect: Background and Motivation}

\subsection{Motivation}
\label{sssec:Why We Need faster Models?}

    Why obsesses on making software analytics faster? Why not just  buy more cloud CPU time?
    Such a ``just throw money at it'' approach might not impress
    researchers like Fisher et al.~\cite{fisher2012interactions} who define ``software analytics'' as a work flow that distills large quantities of low-value
    data down to smaller sets of higher value data. Due to the complexities and computational cost of  some kinds of SE
    analytics, ``the luxuries of interactivity, direct manipulation, and fast system response are gone''~\cite{fisher2012interactions}. They
    characterize modern cloud-based analytics as a throwback to the 1960s-batch processing mainframes where
    jobs are submitted and then analysts wait, wait, wait for results with ``little insight into what's really going
    on behind the scenes, how long it will take, or how much it's going to cost''~\cite{fisher2012interactions}. Fisher et al. document
    the issues seen by 16 industrial data scientists, one of whom remarks ``\underline{{\bf Fast iteration is key}}, but incompatible
    with the say jobs are submitted and processed in the cloud. It's frustrating to wait for hours, only to realize
    you need a slight tweak to your feature set.''.
    
    Fisher's experience matches with our own.   We find that the slower the data mining method, the worse
    the user experience and the fewer the people willing to explore that method. These result is particularly
    acute in research where data miners have to be run many times to (a) explore the range of possible behaviors
    resulting from these methods or (b) generate the statistically significant results that can satisfy peer review.
    To understand the CPU problems with validation, consider the standard validation loop:
    
    {\footnotesize \begin{verbatim}
    1. FOR L = 20 projects DO
    2.   FOR R = 20 times DO # repeats to satisfy central limit theorem
    3      Randomly divide project.data to B = 10 bins;
    4      FOR i = 1..B DO
    5        test = bin[i]
    6        train = project.data - test
    7.       FOR F = 5 different data filters DO
    8.         train = filter(train) # e.g. over-sample rare classes
    9.         model = leaner(train)
    10.        print report(apply(model,test))
    \end{verbatim}
    }
    Note the problem with this loop- it must call a data miner (at line~9) 
    $L * R * B * F = 20,000$ times. While
    some data miners are very fast (e.g. Naive Bayes), some are not (e.g. deep learning). Worse, several ``local
    learning'' results~\cite{menzieslocal,menzies2011local,xia2008local,menzies2013local} report that software analytics results are specific to the data set being processed-
    which means that analysts may need to rerun the above loop anytime new data comes to hand.
    
    Note that the above problem is not solvable by (1)~waiting for faster CPUs or (2)~parallelization. We can no
    longer rely on Moore's Law~\cite{moore1998cramming} to double our computational power very 18 months. Power consumption
    and heat dissipation issues effectively block further exponential increases to CPU clock frequencies~\cite{moore1998cramming}. As
    to parallelization, that would require the kinds of environments that Fisher et al. discuss; i.e. environments
    where it is frustrating to wait for hours, only to realize you need a slight tweak to your feature setting. 
     
    Accordingly, in our research, whenever we have a slow and competent result, we explore methods to make that result faster. The rest of this paper offers a case study where local learner significantly speed up than deep learning.
\subsection{Deep Learning}\label{sec:dl}
    Deep learning is a type of machine learning algorithm based on multiple layers of neural networks, where each layer is created with multiple neurons. These layers are interconnected via weights, which are tuned as the model trains. These connections and weights are very specific to models and its performance. According to LeCun et al.~\cite{lecun2015deep}, deep learning methods are representation-learning methods with multiple levels of representation, obtained by composing simple but non-linear modules that each transforms the representation at one level (starting with the raw input) into a representation at a higher, slightly more abstract level. 
    
    Deep learning has been applied to many areas including
    image processing, natural language processing, genomics etc:
    \bi
    \item Wan and Wang~\cite{wan2014deep},  deep learning tries to resolve the "Semantic Gap" issue in content based image retrieval process~\cite{smeulders2000content}.
    \item
    Deep learning has also been  used to predict DNA-RNA binding proteins~\cite{alipanahi2015predicting} which helps to solve problems faced by less sophisticated methods (particularly the problem of the automatic extraction of meaningful features from raw data).
  \item
    Deep learning has also recently become established its presence in software engineering effort
    estimation~\cite{Choetkiertikul18} and text mining~\cite{choetkiertikul2016deep,mou2016convolutional,white2016deep,white2015toward,yuan2014droid,yang2015deep}. 
    \ei
    Deep learning is a computationally expensive method.  It often takes hours to weeks to train the models. This makes cross validation and stability of model check very expensive if not impractical. For example, the methods of Xu et al. take 14 hours to terminate or GU et al.~\cite{gu2016deep} reported in his paper that his deep learning model took almost 240 hours to train~\cite{gu2016deep}.  Note that if those computations were repeated (say) 20 times for statistically purposes, then those systems would take 11 and 200 days to complete. Worse still, hyperparameter optimization\footnote{
    Hyperparameter optimizers are algorithms that learn the control parameters of a learner.
    For more details on such algorithms, see \S\ref{sssec:Parameter Tuning with DE}.}
    might require 100s to 1000s of repeated runs-- which would take another three to 50 years to terminate\footnote{Note
    that such long runtimes have   been observed in the SE literature. In 2013, a team from UCL needed 15 years of CPU time to complete the hyperparameter optimization study of four software clone detection tools~\cite{Wang13s}.}.
    
For the above reasons, reproducing   deep learning results is a significant problem:
    
    \bi
    \item Most deep learning paper's baseline methods in SE are either not publicly available or too complex to implement~\cite{white2016deep,lam2015combining}.
    \item It is not yet common practice for deep learning researchers to share their implementations and data~\cite{white2016deep,white2015toward,lam2015combining,wang2016automatically,choetkiertikul2016deep,gu2016deep},  where a tiny difference may lead to a huge difference in the results.
    \item Due to the nature of complexity in implementation and unavailability of original implementation, data or environment it is not possible to implement a deep learner for baseline purpose, and this is one of the reasons for SE community and this paper to directly utilizes the results published for comparison~\cite{lam2015combining,fu2017easy}.
    \ei
    Hence, much research (including this paper) is forced to compare their new methods with published numbers in deep learning papers (rather than re-running the rig of the other  researchers.)

    As discussed below,
    local learning is one method for reducing that runtime.

    \subsection{Local Learning}
    When running a data mining algorithm, all the training data can be used to build one training model. Alternatively,
    the training data can be somehow divided into small pieces and one model learned per piece.
    
    Local models have shown promising results in different studies. Many researchers have found building specialized local models for specific regions of the data provides a better overall performance, thus according to the studies instead of trying to find a generalized model we should try to find best models specific to different region of data. 
    For example, Menzies et al.~\cite{menzieslocal} shows that for defect prediction and effort estimation, lessons learned from models build on small part of data set from PROMISE repository were superior to the generalized model build on all the data.
    That said,
    recent studies have suggested for at least for defect prediction, the benefit of local model may be learner specific~\cite{herbold2017global}.
    
    This paper explored local learning
    for a somewhat different perspective.
     The claim made in this paper is not that local models always performs significantly better. Rather we     recommend it since it can lead to significantly faster inference with very little compromise in the performance
     of the learned model. 
     
     More specifically, in this study we observer a three order of magnitude improvement over a prior text mining results by Xu et al~\cite{xu2016predicting}. Hence we recommend local learning since:
    
     \begin{itemize}
        \item It rarely performs worse than learning from all the data.
        \item Sometimes it can lead to better performance~\cite{menzieslocal,menzies2011local,bettenburg2012think,posnett2011ecological}.
        \item It can lead to significantly faster training times.  
    \end{itemize}

    The general framework for local learning is shown as a algorithm in Figure~\ref{fig:local_Learning}. This algorithm uses a cluster based model with the training dataset to find diversity in the dataset. This helps to create clusters with similar data. Now different classification models are accessed and improved via hyperparameter tuning on each of the clusters with local data as training dataset. While accessing the model for performance, the test data is sent through the clustering model to predict its probable cluster by calculating similarity measure, next the test data is classified using the local model built on that cluster.
    
    The rest of this paper explores the local learning framework of Figure~\ref{fig:local_Learning} using 
    the following learning methods:
    \begin{itemize}
    \item
    K-means
    for the clustering;
    \item
    Differential evolution for the fitting;
    \item
    SVM or Kth-nearest neighbor for the classification.
    \end{itemize}
    For details on these methods, see later in this paper.

    \begin{figure}[!t]
    \small
     \begin{lstlisting}[mathescape,linewidth=7.5cm,frame=none,numbers=right]
      def localLearning (data):
        models = []
        predicted = []
        cluster_model = model$_1$(args$_1$) # make clusters
        cluster_model.fit(data.train)
        data.train['cluster'] = cluster_model.labels
        classification_model = model$_2$(args$_2$) # local classifier
        for i in |cluster|
          classification_model.fit(data.train[i]) # fitting model
          models[i] = classification_model
          #end for
        data.test['cluster']= find nearest cluster for test data
        for i in |cluster|
          classification_model = models[i]
          predicted[i]= classification_model.predict(data.test[i])
          #end for
        performance = compare(predicted,data.test['class'])
      return performance    

            \end{lstlisting} 
            \vspace{-0.2cm}
            \caption{Pseudo-code of Local Learning}
            \label{fig:local_Learning} 
            \vspace{-0.3cm}
    \end{figure}

    \subsection{Word Embedding}
    \label{sssec:Word Embedding}
    In the case study of this paper, we explore local learning for text
    mining methods that use word embedding. 
    This is the process of converting words to vectors in order to compare their similarity by comparing cosine similarity between vectors. One method for doing this is a continuous skip-gram model~(word2vec), which is a unsupervised two-layer neural network that converts words into semantic vector representations ~\cite{mikolov2013distributed} and is also used by Fu et al.~\cite{fu2017easy} and Xu et al.~\cite{xu2016predicting} in their paper.
    
    The model learns vector representation of a word~(center word) by predicting surrounding words in a context window($c$) by maximizing the mean of log probability of the surrounding words~($w_{i+j}$), given the center word~($w_{i}$) - 
    
     \begin{equation}
        \frac{1}{n}\sum_{i=1}^{n} \sum_{-c\leq j \leq c, j \neq 0} log p(w_{i+j}|w_i)
    \end{equation}
    
    The  probability $p(w_{i+j}|w_i)$ is a conditional probability defined by a softmax function -
    
    \begin{equation}
        p(w_{i+j}|w_i) = \frac{exp(v_{w_{i+j}}^{'T}v_{w_i})}{\sum_{w=1}^{|W|}exp(v_{w}^{'T}v_{w_i})}
    \end{equation}
    
    Here the $v_{w}$ and $v_{w}^{'T}$ are respectively the input and output vectors of a word $w$ in the neural network, and W is the vocabulary of all words in the word corpus. $p(w_{i+j}|w_i)$ is normalized probability of word $w_{i+j}$ appearing in a specific context for a center word $w_i$. To improve the computation efficiency, Mikolove et al. ~\cite{mikolov2013distributed} proposed
    hierarchical softmax and negative sampling techniques etc which can also be used for creating word embedding models.
    
    This paper uses the word2vec models trained by Fu et al.~\cite{fu2017easy} that converted the Stack Overflow text data into the corresponding vectors. For our experiments, 
    we use    $100,000$ randomly selected  knowledge units tagged with ``java'' from
    Stack Overflow {\it posts} table  (include titles, questions and answers).
    This data was pruned by removing superflous
    HTML tags (while keeping short code snippets in {\it code} tag), then fitted  into the {\it gensim} word2vec module ~\cite{rehurek2010software}.
    This is  a python wrapper over original word2vec package where,  for   word $w_i$ in the post is sent to the trained word2vec model to get the corresponding word vector representation $v_i$. After converting, all the KUs the output vectors are then used for training and testing the models.
    
    \subsection{SVM in Text Mining}
    \label{sssec:SVM in Text Mining}
    Another component  of the case study explored here is support vector machines.
    SVMs are a type of supervised machine learning algorithm which analyzes data using classification or regression analysis. In SVM models the examples are points in space mapped in such a way that separate categories/classes are divided by a clear gap (hyperplane in instance space)~\cite{suykens1999least}.
    SVMs execute by  transforming the original data space to a higher dimensional space where hyperplane between data from different classes is easier to detect.
    
    SVMs are particularly useful for  in text mining can have a very large number of features~\cite{gamon2004sentiment}.  In most cases the document vectors are sparse and linearly separable in some hyper-dimensional space~\cite{joachims1998text}.
    
    \subsection{KNN in Text Mining}
    \label{sssec:KNN in Text Mining}
    If SVM is our most sophisticated classifier, our simplest is K-Nearest Neighbor~\cite{zhang2007ml}. KNN is a non-parametric ~\cite{goldberger2005neighbourhood} method used for classification and regression problems. Here {\em k} is the input and refers to the number of closest examples that the model will look for among the training data in the feature space. The output the model gives represents the class to which the test data belongs to and it depends on the majority vote of its k-nearest neighbor ~\cite{mihalcea2006corpus}.
  
    
    \subsection{KMeans Clustering}
    \label{sssec:K-Means Clustering}
    One way to reduce the computational cost of KNN is, before running a classifier,  group data into similar sets of clusters. KMeans is an unsupervised machine learning algorithm that is used for clustering data ~\cite{jain2010data}. In KMeans, {\em k} is an input that refers to the number of clusters the data set should be divided into. The algorithm initializes {\em k} number of centroids from the data and labels each as a cluster. For each data point it checks which centroid it is closest to, and assigns it to that cluster. After one pass to the data, centroids are recalculated and the process repeats until cluster stability is achieved. 
    This experiments uses the scikit-learn module   sklearn.cluster.KMeans~\cite{pedregosa2011scikit}.
    
    \begin{figure}[!t]
    \small
     \begin{lstlisting}[mathescape,linewidth=7.5cm,frame=none,numbers=right]
      def GAP( tData, nrefs=3, cMax = 15): # default settings
        gaps = []
        results = create empty dataframe
        for gap_index, k in enumerate(range(1, cMax)):
          refDisps = array for inertia
          for i in range(nrefs):
              rRef = random(tData) #reference data
              clf = KMeans(k)
              clf.fit(rRef)
              refDisps[i] = model inertia
          #end for
          clf = KMeans(k)
          clf.fit(tData)
          orgDisp = model inertia with k cluster
          refDispMean = mean(refDisps)
          dispDiff = refDispMean - orgDisp
          gap = log(dispDiff)
          # append the gap value in the list 
          results = results.append('clusterCount': k, 'gap': gap) 
          gaps[gap_index] = gap
      return gaps.argmax() # get cluster size with max gap
            \end{lstlisting} 
            \vspace{-0.2cm}
            \caption{Pseudo-code of GAP Statistics}
            \label{fig:GAP_pseudocode} 
            \vspace{-0.3cm}
    \end{figure}
     \begin{table*}[!t]
    \small
        \centering
       \begin{tabular}{c|c|c|c}
         \rowcolor{darkgray} \textcolor{white}{\textbf{SVM}} & \textcolor{white}{\textbf{Default Parameter}}  & \textcolor{white}{\textbf{Tuning Range }} & \textcolor{white}{\textbf{Description}} \\
            C & 1.0 & [1,50] & Penalty parameter of  error term \\ 
            \rowcolor{lightgray} Kernal & `rbf' & [`liner', `poly', `rbf', `sigmoid'] & Specify the kernel type to be used in the algorithms \\ 
            gamma & 1/n\_features & [0,1] & Kernel coecient for `rbf', `poly' and `sigmoid' \\ 
             \rowcolor{lightgray}coef0 & 0 & [0,1] & Independent term in kernel function. It is only used in `poly' and `sigmoid' \\
            \multicolumn{1}{c}{~}\\
            \rowcolor{darkgray} \textcolor{white}{\textbf{KNN}} & \textcolor{white}{\textbf{Default Parameter}} & \textcolor{white}{\textbf{Tuning Range }} & \textcolor{white}{\textbf{Description}} \\ 
            n\_neighbors & 5 & [2,10] & Number of neighbors \\ 
            \rowcolor{lightgray} weights & `uniform' & [`uniform', `distance'] & weight function used for predictions 
        \end{tabular}    
        \caption{`Tuning Range' of Parameters for SVM and KNN}
        \label{tab:DE_Parameters}
    \end{table*}
    
    Picking an appropriate {\em k} value for KMeans is a challenge. If {\em k} is too small,  models will over-fit and  fails to capture any larger patterns in the data. On the other hand, if using a large {\em k} will increase the variability and hence the level of uncertainty within each cluster. accordingly, this study uses the {\em GAP statistics} to determine the optimal number of {\em k} or centroids for KMeans ~\cite{mohajer2011comparison,tibshirani2001estimating}. The GAP statistic looks at the difference between the dispersion of the clustered data, and the dispersion of a null reference distribution, for increasing {\em k} values. It finds the largest {\em k} where the gap is bigger than the next gap minus a value that accounts for simulation error.   Figure~\ref{fig:GAP_pseudocode} describes the GAP statistic
    computation~\cite{tibshirani2001estimating}.
    
    \subsection{Parameter Tuning with DE}
    \label{sssec:Parameter Tuning with DE}
    All learners come with ``magic parameters'' that control their performance.
    For example, with SVM, there are several parameter that control the SVM kernel.
    One way to select those parameters is to use hyperparameter optimization via
    algorithms like  
    Differential Evolution. DE is a stochastic population-based optimization algorithm~\cite{storn1997differential}. DE starts with a frontier of randomly
    generated candidate solutions.
    For example, when exploring tuning, each member of the frontier would be a different possible set of control settings for (say) an SVM or KNN.
    
    After initializing this frontier, a new candidate solution is generated by extrapolating
    by some factor $f$ between other items on the frontier. Such extrapolations are
    performed for all attributes at probability {\em cf}.
    If the candidate is better than one item of the frontier, then the candidate replaces the frontier item. The search then repeats for the remaining frontier items.
    For the definition of ``better``, this study uses the same performance measures as Fu et al.; i.e. 
   ``better'' means maximizing  the objective score of the model based F1 Score.

    This process is repeated for {\em lives} number of repeated traversals of the frontier.
        For full details of DE,  see fFgure~\ref{fig:pseudo_DE}. As per Storn's advice~\cite{storn1997differential} we use \[f=0.75, \mathit{cf}=0.3,
    \mathit{lives}=60\]

    \begin{figure}[!t]
    \small 
    \begin{lstlisting}[mathescape,linewidth=7.5cm,frame=none,numbers=right ]
      def DE( n=10, cf=0.3, f=0.7):  # default settings
        frontier = sets of guesses (n=10)
        best = frontier.1 # any value at all
        lives = 1
        while(lives$--$ > 0): 
          tmp = empty
          for i = 1 to $|$frontier$|$: # size of frontier
             old = frontier$_i$
             x,y,z = any three from frontier, picked at random
             new= copy(old)  
             for j = 1 to $|$new$|$: # for all attributes
               if rand() < cf    # at probability cf...
                  new.j = $x.j + f*(z.j - y.j)$  # ...change item j
             # end for
             new  = new if better(new,old) else old
             tmp$_i$ = new 
             if better(new,best) then
                best = new
                lives++ # enable one more generation
             end                  
          # end for
         frontier = tmp
        # end while
        return best
    \end{lstlisting} 
    \caption{Tuner Procedure - as mentioned in Fu et al.\textquotesingle s paper. It is based on Storn\textquotesingle s DE optimizer.}
    \label{fig:pseudo_DE} 
    \vspace{-0.3cm}
    \end{figure}

    For SVM, this study uses the SVM module from Scikit-learn~\cite{pedregosa2011scikit}, a Python package for machine learning, where the following parameters shown in Table~\ref{tab:DE_Parameters} are selected for tuning:
    \bi
    \item
    Parameter $C$ is to set the amount of regularization, which controls the trade-off between the errors on training data and the model complexity. A small value for $C$ will generate a simple model with more training errors, while a large value will lead to a complicated model with fewer errors. 
    \item 
    {\em gamma} defines how far the influence of a single training example reaches, with low values meaning 'far' and high values meaning `close'.
    \item
    {\em coef0} is an independent parameter used in sigmod and polynomial kernel function to scale the data.
    \ei
    Similarly KNN uses different parameters to control how it learns and how it predicts. This study uses KNN module from Scikit-learn, and the parameters in   Table~\ref{tab:DE_Parameters}. For KNN:
    \bi
    \item Parameter $n\_neighbors$ is number of neighbors to be used for query to check and   classification using a  majority vote ~\cite{guo2003knn}. 
    \item
    $weight$ is another parameter which is tuned as part of hyperparameter tuning. If set to `uniform', then   all the points in each neighborhood are weighted equally. If set
    to `distance', then the weights are inverse of their distance from the new example (so farther away the point, the less weight  it has in deciding the class).
    \ei

\section{Experimental Design}
\label{sect: Experimental Design}
    \subsection{Data}
    \label{sssec:Data}
    This experiment uses the same training and testing dataset as Xu et al. ~\cite{xu2016predicting} and Fu et al. ~\cite{fu2017easy}. For the reasons discussed
    in Section~\ref{sec:dl}, this study compares our results to those reported by  Xu et al.  
    
    Our dataset includes 6,400 training examples and 1,600 testing examples. Each class is equally represented in both the training and test datasets, with 1600 of each class in the training dataset and 400 of each class in the test dataset, so no handling of class imbalance is necessary.
    
    Both test and training data are in Pandas dataframe ~\cite{mckinney2011pandas} format, which includes a post Id, related post id, and a link type which is determined by a score between the 2 posts. The link between 2 posts can be of 4 types depending on the score between the sentences as per Table~\ref{tab:data_classl}.
    
    \begin{table}[h!]
        \centering
        \begin{tabular}{c|c|c}
            \rowcolor{darkgray} \textcolor{white}{\textbf{Scores}} & \textcolor{white}{\textbf{Class ID}} & \textcolor{white}{\textbf{Type}} \\
            1.0 & 1 & duplicate  \\
            \rowcolor{lightgray} 0.8 & 2 & direct link \\
            0<x<0.8 & 3 & indirect link \\
            \rowcolor{lightgray} 0.00 & 4 & isolated \\
        \end{tabular}
        \caption{Classification of data}
        \label{tab:data_classl}
    \end{table}
        \begin{figure*}
        \centering
        \tcbox[sharp corners, boxsep=1mm, boxrule=0.5mm, 
            colframe=blue!30!black, colback=white]{\includegraphics[width=0.8\textwidth]{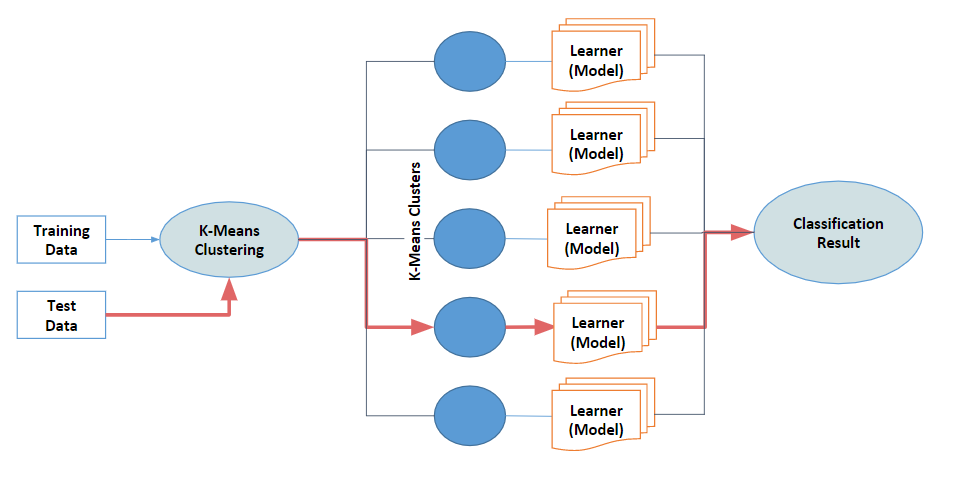}}
        \caption{Model Architecture}
        \label{fig:our_model}
    \end{figure*}

    For finding word embedding ~\cite{mikolov2013distributed} of Stack Overflow ~\cite{barua2014developers} post data, this paper uses the word2vec model trained by Fu et al. and described in Section \ref{sssec:Word Embedding}.
     The final data that     passed to our classifiers is similar to Table~\ref{tab:data_format}.

    \begin{table}[]
        \centering
        \label{my-label}
            \begin{tabular}{l|l|l|l|l|l|l}
            \rowcolor[HTML]{9B9B9B} 
            {\color[HTML]{FFFFFF} \textbf{ID}} & {\color[HTML]{FFFFFF} \textbf{\begin{tabular}[c]{@{}l@{}}Post\\  Id\end{tabular}}} & {\color[HTML]{FFFFFF} \textbf{\begin{tabular}[c]{@{}l@{}}Related \\ Post Id\end{tabular}}} & {\color[HTML]{FFFFFF} \textbf{\begin{tabular}[c]{@{}l@{}}Link \\ Type \\ Id\end{tabular}}} & {\color[HTML]{FFFFFF} \textbf{\begin{tabular}[c]{@{}l@{}}Post \\ Id Vec\end{tabular}}} & {\color[HTML]{FFFFFF} \textbf{\begin{tabular}[c]{@{}l@{}}Related \\ Post Id \\ Vec\end{tabular}}} & {\color[HTML]{FFFFFF} \textbf{Output}} \\
            0 & 283 & 297 & 1 & {[}...{]} & {[}...{]} & {[}...{]} \\
            \rowcolor{lightgray} 1 & 56 & 68 & 2 & {[}...{]} & {[}...{]} & {[}...{]} \\
            2 & 5 & 16 & 3 & {[}...{]} & {[}...{]} & {[}...{]} \\
            \rowcolor{lightgray} 3 & 9083 & 6841 & 4 & {[}...{]} & {[}...{]} & {[}...{]}
            \end{tabular}
        \caption{Training/Test data in Pandas.dataframe format.}
        \label{tab:data_format}
    \end{table}

    \subsection{Method}
    \label{sssec:Method}
    Training a models and hyper-parameter tuning technique can take much time, depending on the complexity of the training dataset. This makes it harder to perform cross validation or repeatability of prior results. This study check if dividing up the data into small clusters and then train and tune models within each cluster, reduces
    the overall learner's training time. In order to do this the data is clustered first, then a model is built for each cluster. This process is shown   Figure~\ref{fig:our_model}.
    
    For the clustering algorithm KMeans algorithm from Scikit-learn has been used. For the parameters, this study used the k-means++ algorithm ~\cite{arthur2007k} for the initialization of cluster centroids. In order to choose {\em k}, the number of clusters, the GAP statistic method, discussed in Figure~\ref{fig:GAP_pseudocode} has been utilized.
    
    For each cluster, a learner is built that is tuned specifically for that cluster. This study looks at two different learners, SVM and KNN. As discussed above, to tune the learners, DE is being used, specifically Fu et al.\textquotesingle s DE implementation. The study uses F1 Score ~\cite{sokolova2006beyond} to evaluate the intermediate models in DE, as F1 Score is calculated as the trade-off between precision and recall. This will help us to get models which have both good precision and recall.

    To use the model, the test data is first sent to KMeans to find the cluster which it should belong to by calculating vector distance from all cluster's centroid and returning the cluster with minimum distance. Then the model predict the class using that cluster\textquotesingle s learner.
    
    In this study a 10-fold cross validation ~\cite{kohavi1995study} has been used, which was repeated 10 times for the training data. Thus, the results are the mean of 100 models. Each learner (SVM or KNN) on each cluster has been trained on 90\% of the data from that cluster and tuned on rest of the 10\% and then tested on the untouched test data set.

    \subsection{Performance Criteria}
    \label{sssec:performance_criteria}
    For evaluating the described model, this study collects and present the same metrics for performance evaluation as Xu et al. and Fu et al., in order to compare results. These metrics are precision, recall and F1 Score. This multiclass classification problem have 4 classes denoted as Duplicate(C1), Direct Link(C2), Indirect Link(C3) and Isolated(C4). The result presents class-wise metrics as well as the mean for the whole model.
    
    \begin{table}[h!]
        \centering
        \small
        \resizebox{0.5\linewidth}{!}{\begin{tabular}{@{}cc|c|c|c|c|l@{}}
            \cline{3-6}
            & & \multicolumn{4}{ c| }{Classified as} \\ \cline{3-6}
            & & $C_1$ & $C_2$ & $C_3$ & $C_4$ \\ \cline{1-6}
            \multicolumn{1}{ |c  }{{\rotatebox[origin=c]{90}{Actual}}} &
            \multicolumn{1}{ |c| }{$C_1$} & \cellcolor{lightgray}$C_{11}$ & $C_{12}$  & $C_{13}$ & $C_{14}$  & \\ \cline{2-6}
            \multicolumn{1}{ |c  }{}                        &
            \multicolumn{1}{ |c| }{$C_2$} & $C_{21}$ & \cellcolor{lightgray}$C_{22}$ & $C_{23}$ & $C_{24}$ &  \\ \cline{2-6}
            \multicolumn{1}{ |c  }{}                        &
            \multicolumn{1}{ |c| }{$C_3$} & $C_{31}$ & $C_{32}$ & \cellcolor{lightgray}$C_{33}$ & $C_{34}$ & \\ \cline{2-6}
            \multicolumn{1}{ |c  }{}                        &
            \multicolumn{1}{ |c| }{$C_4$} & $C_{41}$ & $C_{42}$ & $C_{43}$ & \cellcolor{lightgray}$C_{44}$ & \\ \cline{1-6}
        \end{tabular}}
        \caption{Confusion Matrix}
        \label{tab:confusion_matrix}
    \end{table}
    
    From the confusion matrix describe in Table~\ref{tab:confusion_matrix}, it can be observed that the correct predictions are the one with label {\em Cii} for a class {\em Ci}. And from this we will be able to calculate our evaluation matrix. Using
    this nomenclature, the evaluation metric {\em F1 Score} can be defined as follows:

    \begin{equation}
        \mathit{precision} = \frac{C_{ii}}{\sum\limits_{i}C_{ij}}
    \end{equation}
        \begin{equation}
        \mathit{recall} = \frac{C_{ii}}{\sum\limits_{i}C_{ji}}
    \end{equation}
        \begin{equation}
        \mathit{F1} = \frac{2*\mathit{recall}*\mathit{precision}}{(\mathit{precision} + \mathit{recall})}
    \end{equation}

    \subsection{Statistical Analysis}
    \label{sssec:Statistical Analysis}
    In order to compare results of the local models with other models, there are two useful tests: significance tests ~\cite{bentler1980significance} and effect size tests ~\cite{rosenthal1994parametric}~\cite{chen2002correlation}.
   \bi
   \item
  Significance tests assess if  results are distinct enough to be considered different;
  \item
  Effect size tells us whether that difference is large enough to be  interesting. 
  \ei
  After Wu et al.,  this study uses the Scott-Knott test, which ranks treatments using a recursive bi-clustering algorithm. At each level, treatments are split where expected value of the treatments has most changed from before. Results in each rank are considered the same according to both significance and effect size tests. As per the recommendations of Wu et al., the Scott-Knott test uses the nonparametric bootstrap ~\cite{efron1982jackknife} method and Cliffs' Delta. Note that these two tests are also used and endorsed by other SE
  researchers~\cite{ghotra2015revisiting}.
  \begin{table}[!b]
        \centering
        \begin{tabular}{c|c}
            \rowcolor{darkgray} \textcolor{white}{\textbf{Class}} & \textcolor{white}{\textbf{F1 Score Mean}} \\
            \rowcolor{darkgray} & \begin{tabular}{c|c}
                 \textcolor{white}{\textbf{Our DE\_SVM   }}  &  \textcolor{white}{\textbf{Fu's DE\_SVM}} \\
             \end{tabular}\\
            Duplicate & \begin{tabular}{c|c}
                92  & 88 \\
             \end{tabular}\\
            \rowcolor{lightgray} 
            Direct link & \begin{tabular}{c|c}
                91  & 84 \\
             \end{tabular}\\
            Indirect link & \begin{tabular}{c|c}
                98  & 97 \\
             \end{tabular}\\
            \rowcolor{lightgray} 
            Isolated & \begin{tabular}{c|c}
                93  & 91 \\
             \end{tabular}\\
            Overall & \begin{tabular}{c|c}
                93  & 90 \\
             \end{tabular}\\
        \end{tabular}
        \caption{Comparison of all performance measure between Our DE\_SVM and Fu et al.\textquotesingle s DE\_SVM }
        \label{tab:performance Measure comp}
    \end{table}
\section{Results}
\label{sect:Result}
    
    \textbf{RQ1}:  {\em Can we reproduce Fu et al.\textquotesingle s results for tuning SVM with differential evolution (DE)?}

    This study uses same differential evolution with SVM for both global and local models. Thus to compare with Fu's DE with SVM as global model, the first task as part of this experiment was to recreate Fu et al.\textquotesingle s work so that this study have a baseline to measure against.
  Hence, this research question is a  ``sanity check'' that must be passed before moving on to the other, more interesting research questions.

    The study uses the same SVM from Scikit-learn with the parameters tuned as mentioned in Table~\ref{tab:DE_Parameters}. Here the training time of the DE+SVM model is also compared with Fu et al.\textquotesingle s model. Table~\ref{tab:performance Measure comp}   shows the class by class comparison for all the performance measure this study is using.
    
    From Table~\ref{tab:performance Measure comp}  it can be seen that our results  with SVM with DE for hyperparameter tuning ~\cite{duan2003evaluation}~\cite{fu2017easy} similar to the results of Fu et al. It can be observed from this figure that for most of the cases apart from class 3, the model has performed a little better, but the delta between the performance is very small. 
    
    Hence the answer to our  RQ1, is that this study has successfully implemented Fu et al.\textquotesingle s SVM. Hence, we can move to more interesting questions.

    \begin{figure}[!t]
        \centering
        \includegraphics[width=\linewidth]{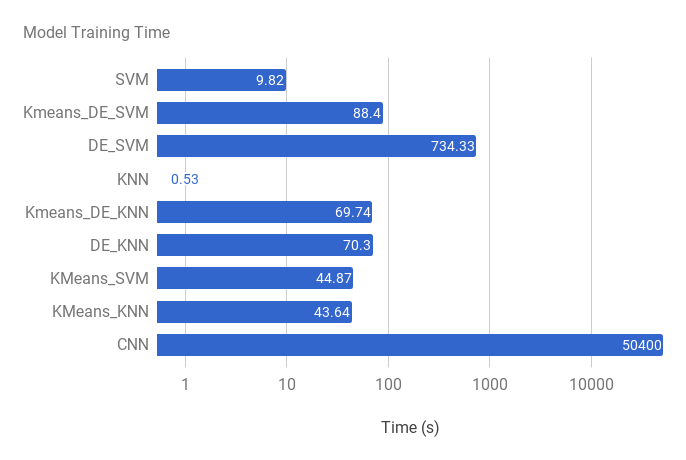}
        \caption{Training time comparison between models in Log Scale. In the above,
        the Xu et al. 2016 results are labelled ``CNN'' and the Fu et al. 2017
        results are labelled ``DE\_SVM''.}
        \label{fig:time}
    \end{figure}
    

   \textbf{RQ2}: {\em  How do the local models compare with global models in both tuned and untuned versions in terms model training time?}

    For RQ2, this experiment built one model for each clusters using either normal or tuned versions of 
     SVM or  KNN (where tuning was performed with DE):
      For the default SVM and KNN the experiment uses the default parameters, described in Table~\ref{tab:learners}. 
  
     As discussed above, this study have used the GAP statistic~\cite{mohajer2011comparison}~\cite{tibshirani2001estimating} for finding the best number of clusters, using minimum and maximum number of clusters as 3 and 15, respectively. As part of the experiment we learned that 13 clusters achieves  best results 
     (measured as per the GAP statistic).
     
    This study measures the time taken for this model to train which includes time taken by GAP statistic, KMeans training time, and SVM/KNN with DE training time.

    Figure~\ref{fig:time} compare the model training time in log scale  of all models with the results
    from XU et al.\textquotesingle s CNN approach.
    Its apparent from the figure~\ref{fig:time} that for this domain KNN and SVM has the fastest runtimes. That said, as describe below, we cannot   recommend these
    methods since, as shown below, they achieve poor F1 Scores.
    
    The figure also shows, when we cluster data and then create local models on each cluster it achieves fast 
    training times.  For example,
    in the case of KMeans\_DE\_SVM, we achieve an eight-fold  speed improvement from using DE\_SVM. 
    Further, we see a speed improvement of over 570 faster training time from XU\textquotesingle s CNN\footnote{
    Technical aside: most of  our times  come from a   single core, single thread implementation.
    That said,  just for completeness, we have trained our   $k$ clusters over a standard 8 core laptop. In that multi-threaded implementation,  our training times
    are 965 times faster than  XU\textquotesingle s CNN.  }.
     

    \textbf{RQ3}: {\em  How does the performance of local models compare with global models and state-of-the-art Deep Learner when used with SVM and KNN?} 
    


    \begin{figure}
        \centering
        \includegraphics[width=\linewidth]{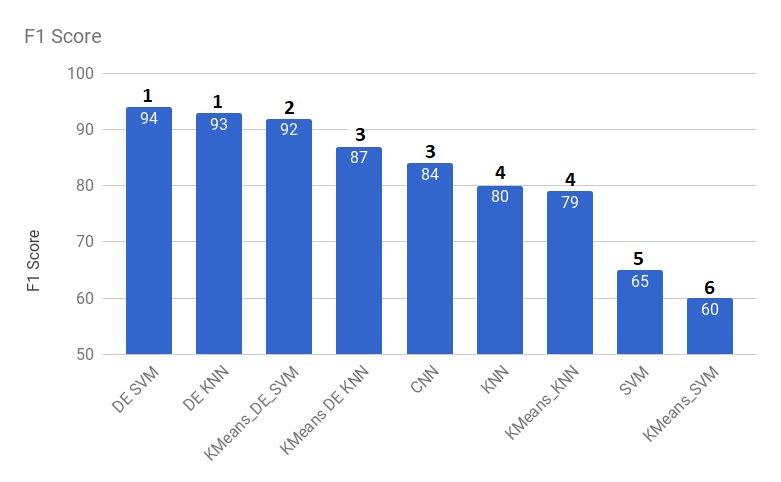}
        \caption{F1 Score comparison for all learners. Numbers above each bar are ranks learned from the Scott-Knot statistical test. Clusters have different ranks if they are determined to be different
        by both a significance test (bootstrapping) and an effect size test (Cliffs' Delta). For example, the first two bars have the same rank while the third bar has a different rank. In the above,
        the Xu et al. 2016 results are labelled ``CNN'' and the Fu et al. 2017
        results are labelled ``DE\_SVM''.}
        \label{fig:F1-Score}
    \end{figure}
    
    The final part of our research question was to check if the local models performance is comparable to Fu et al.'s DE\_SVM and the XU's state of the art CNN. To evaluate the performance of the models this study compares F1 performance measures described in Section~\ref{sssec:performance_criteria}. As mentioned in the section, a 10 fold * 10 repeat cross validation was  performed, so all the results are mean of 100 models created. 
    
    Figure~\ref{fig:F1-Score} shows our  F1 Score results (mean result across
    all 4 class of Table~\ref{tab:performance Measure comp}). The numbers on top of each bar show the results of statistical tests. Bars with the same rank are statistically indistinguishable.
       Note that these results should be discussed with respect to the runtime results shown above:
       
       \bi
       
        \item
      In  Figure~\ref{fig:time}, we saw that  CNN was our  slowest learning method.
       We would excuse this slowness and endorse its use if the performance scores for this extremely slow method  was outstandingly high. This is {\em not} the case: as seen
       Figure~\ref{fig:F1-Score}, 
        CNN is beaten by many of the treatments shown here.
       \item
       In  Figure~\ref{fig:time},  DE\_SVM was our second slowest method. Again, we would endorse this approach if nothing else comes close to it in terms of performance.
       While KMeans\_DE\_SVM is significantly different (as shown by our Scott-Knott results), the median performance delta is very small indeed
       (median F1 Score scores of 94 vs 92). This is an important pragmatic
       consideration since, as shown in Figure~\ref{fig:time}, the learning time of 
       KMeans\_DE\_SVM  was  743/88 = 840\% faster than that of  DE\_SVM
       \ei
       
      \noindent In summary, we recommend one of our  local learning method (KMeans\_DE\_SVM) since:
       \bi
       \item It is 840\%   faster than the prior state-of-the-art results (Fu et al.'s 2017 DE\_SVM method);
       \item
       On a single core machine, it is 570 times faster
       than the prior state-of-the-art before that (Xu et al.'s 2016 CNN method);
       \item
       On a standard laptop with  8 cores, KMeans\_DE\_SVM  runs 965 times faster than Xu et al.'s CNN method;
       \item
       Our local learner performs better than Xu et al. and only a tiny fraction worse than Fu et al.   Given these small performance deltas,
            from a pragmatic engineering perspective,
              we find it hard to justify the extra computational cost of
               DE\_SVM over  KMeans\_DE\_SVM. 
       \ei

\section{THREATS TO VALIDITY}
\label{sect:THREATS TO VALIDITY}
    As with any empirical study, biases can affect the final results. Therefore, any conclusions made from this work must be considered with the following issues in mind:
    \bi
        \item Sampling bias: threatens any classification experiment; i.e., what matters there may not be true here. For example, the data sets used here is a Stack Overflow dataset and were supplied by one individual. Although this study uses multiple word2vec models to validate with a 10-fold * 10 repeat validations. The text data is of similar format of question pairs.
        
        \item Learner bias: For building the model for finding semantic relatedness of question pars in this study, we elected to use Support Vector Machine and K-Nearest Neighbor for classification and KMeans for clustering. This study chose these methods because its results were comparable to the more complicated algorithms and has been successful in text classification field. Classification and clustering is a large and active field and any single study can only use a small subset of the known algorithms. 
        
        \item Evaluation bias: This paper uses precision, recall and F1 Score as performance measures. Other methods like Concordant - Discordant ratio~\cite{goodman1972measures}, Gini Coefficient~\cite{neslin2006defection}, Kolmogorov Smirnov chart~\cite{guo2011case} that are used for this purpose can also be used for performance evolution in future studies.

        \item Input bias: For the localization algorithm, this study randomly selects input values for a range to determine the number of clusters, also for hyperparameter tuning using DE, a subset of parameters has been selected for tuning and their range is either the whole range that the parameter accepts or a range that is selected for the study.
    \ei

\section{CONCLUSION}
\label{sect:CONCLUSION}
   This paper investigates the value of local learning from the perspective of  reducing the   training time compared to 
    methods for  predicting knowledge unit's relatedness on Stack Overflow. This study   clustered the data first;  then build local models on each cluster separately; then use the clustering algorithms with the local classification algorithm together to predict. As shown above:
    
    \bi
        \item Clustering the data first and then building local models on those subsets of data shows significant reduction in runtime.
        \item Using KMeans on the Word Embedding model first to cluster the data into smaller subsets and then running SVM and KNN with their DE versions showed runtime reduction as large as 570 times for KMeans\_DE\_SVM version with CNN and almost 8 times improvement than its global DE\_SVM version with sequential run. And with parallel run the performance improvement for KMeans\_DE\_SVM version with CNN is 965x and almost 14 times with its global counter part.
        \item The performance in term of precision, recall and F-score is almost similar to its global counter part and sometime better then the XU's CNN model.
    \ei
    

    For future work, we suggest trying several variations of this experiment - 
    
    \begin{itemize}
        \item This study tunes the local models separately, tuning all the local models along with the clustering algorithm together might produce some interesting results.
        \item Trying to find correlation between number of cluster and training time and performance might give us effect of number of clusters on the prediction model.
    \end{itemize}

\bibliographystyle{ACM-Reference-Format}
\bibliography{700faster} 

\end{document}